\documentclass[aps,pra,twocolumn,showpacs,superscriptaddress]{revtex4}
\usepackage{bm,amsmath,amssymb,latexsym,graphicx,enumerate}

\usepackage{epsfig, graphicx, float}
\usepackage{xcolor}

\newcommand{\be}{\begin{equation}}
\newcommand{\ee}{\end{equation}}
\newcommand{\pa}{\partial}
\newcommand{\RR}{\mathbb{R}}

\begin{document}

\title{Stopping a reaction-diffusion front}
\author{Jean-Guy Caputo}
\affiliation{Laboratoire de Math\'ematiques, INSA de Rouen - B.P. 8, Avenue de l'Universit\'e, 76801 Saint-Etienne du Rouvray, France.  E-mail: caputo@insa-rouen.fr}
\author{Gustavo Cruz-Pacheco}
\affiliation{Depto. Matem\'{a}ticas y Mec\'{a}nica,
I.I.M.A.S.-U.N.A.M., Apdo. Postal 20--726, 01000 M\'{e}xico D.F., M\'{e}xico}
\author{Beno\^it Sarels}
\affiliation{Sorbonne Universit\'e, CNRS, Universit\'e de Paris,
Laboratoire Jacques-Louis Lions (LJLL), F-75005 Paris, France}

\begin{abstract}
We revisit the problem of pinning a reaction-diffusion front by a defect, in 
particular by a reaction-free region. Using collective variables for the front
and numerical simulations, we compare the behaviors of a bistable and monostable
front. A bistable front can be pinned as confirmed by a pinning criterion,
the analysis of the time independant problem and simulations. Conversely,
a monostable front can never be pinned, it gives rise to a secondary pulse
past the defect and we calculate the time this pulse takes to appear. These 
radically
different behaviors of bistable and monostable fronts raise issues for
modelers in particular areas of biology, as for example, the study of 
tumor growth in the presence of different tissues.
\end{abstract}
\maketitle

\section{Introduction}

Reaction-diffusion equations are general models for active chemical
reactions like combustion \cite{zeldovich}, see the book by Alwyn Scott \cite{scott} 
for examples, or the review \cite{xin00}.
In biology, while these equations are ubiquitous in population dynamics or population genetics,
they are also used in a variety of problems such as nerve impulse propagation in an axon
or growth of a cancer tumor \cite{cancer},
see the general book by Murray \cite{murray}.
These equations have a number of stationary solutions. 
In the physics context, the models have three stationary
solutions, two stable and one unstable, they are termed {\it bistable}.
In biology, the bistable model can be used to describe the growth of a population.
Still, the main model remains the one that describes the wave of advance of an advantageous gene like in the seminal paper \cite{fisher}.
It only has two stationary states, one stable 
and another unstable; it is called {\it monostable}.
The normal forms of these two different
reaction terms are cubic for the bistable and quadratic for the monostable.
Mostly, the study of these two nonlinearities gives the qualitative picture for
almost all other nonlinearities.

Important dynamical solutions
are the fronts (in 1D) that connect two such stationary solutions. In
2D and 3D, assuming radial symmetry, fronts describe the 
interface of "blobs" where the inside has one state and the outside 
another. Fronts have a speed that is proportional to the square
root of the diffusion times the reaction rate. Many studies 
dealt with the stability of such fronts, see \cite{scott}
for references. Assuming a front is stable, important parameters 
are its position and width i.e.
the spatial extension of the transition region separating the two stationary
states. These fronts have exact forms for the bistable \cite{scott} 
and monostable \cite{ablowitz} nonlinearities. Using these solutions
as Rayleigh-Ritz type ansatzes, we and other authors have derived 
ordinary differential equations giving the evolution of 
the position and width -termed
collective coordinates- whose solutions
match remarkably well the full dynamics, see \cite{cs1,susanto} for
the bistable case and \cite{bcp14} for the monostable case.

For many applications, it is useful to understand how fronts respond
to perturbations of the environment. Such perturbations can be temporal such
as the action of therapy on a cancer cell \cite{cancer2}. They can also
be spatial like a defect in a burning candle. Such a geographic defect
can act on scales larger or smaller than the front width. In the latter
case, the front adiabatically adapts to the defect and changes its width
and speed accordingly, see for example our study \cite{cs1} on bistable
fronts. When the defect is narrow compared to the front, a bistable front
can be pinned by the defect \cite{cs1}. This effect can also be seen in 
waveguides whose transverse width abruptly changes, see the works
\cite{bbc16} and \cite{thesebouhours}.

In this article, we revisit the issue of front pinning comparing
bistable and monostable fronts. As expected, wide defects cannot pin any of
the two types of fronts. Narrow defects will pin a bistable front, see the
numerics and analysis of \cite{cs1}. To understand
further the phenomenon, we consider as defect a no reaction region
of a given extension. Using collective variables for the front, we find
a pinning criterion which indicates that static bistable fronts exist 
in such a region if it is large enough. This approach is confirmed
by analyzing the time independant problem and the solution of the
1D partial differential equation.
Monostable fronts behave very differently, they can never be pinned.
Numerical simulations of monostable
fronts appear to stop the front but the wave continues to propagate 
and gives rise to a secondary pulse, past the defect.
We calculated the time of delay, i.e. the difference between the 
arrival at the defect and the appearance of the secondary pulse, when
its maximum reaches 0.5. This delay time scales approximately 
linearly with the extension of the zero reaction zone, in agreement
with a simple calculation based on the diffusion kernel and the 
instability rate of the zero stationary solution. \\
The article is organized as follows. The bistable and monostable
models and their exact solutions are recalled in section 2. Section 3
presents and discusses the collective variable differential equations 
for the front position and width. Sections 4 and 5 detail the front motion
in a no reaction zone for the bistable and monostable models respectively.
Section 6 concludes the article.

\section{The model}
In the following we are concerned with the equation
\be\label{rd}
\left\{
\begin{aligned}
u_t &= u_{xx}  +  s(x)~R(u) \\
u(x, 0) &= u_0(x) \\
\end{aligned}
\right.
\ee
for $(x, t) \in \RR \times \RR_+^*$ and $u_0$ a given function with sufficient regularity.
We will study the two canonical types of non-linearities:
the monostable one $R(u) = u(1-u)$
and the bistable one $R(u) = u(1-u)(u-a)$ for some $0<a<1$.
To account for the variable growth rate of the quantity $u$ (chemical density, population density...),
we make use of a reactivity $s$ that is space-dependent
and remains positive over the considered range: $s(x)>0$ for every real $x$.
Let us first recall the theory when $s$ is constant.
We recall that in the former case, the state $u^*=1$ is a stable equilibrium of the associated ordinary differential equation
while $u^*=0$ is an unstable equilibrium.
In the latter, the states $u^*=0,1$ are stable equilibria
while $u^*=a$ is unstable.

To carry out our analysis in the next sections,
we will make extensive use of the fact that exact solutions of equation \eqref{rd} are known when $s$ is constant.
In the bistable case, all traveling wave solutions are fronts connecting the two stable equilibria.
We choose to consider in the rest of the paper only fronts going from $1$ at $-\infty$ to $0$ at $+\infty$.
Then all traveling wave solutions - up to a translation - are of the form
\be \label{exactbi}
u(x,t) = \tilde{U}_{bi}(x-ct) = \frac{1}{1 + \exp { \left(\sqrt{\frac{s}{2}} (x-ct) \right)}} ,
\ee
where the speed $c$ is known \cite{scott} and related to the parameter $a$ \emph{via} the formula
\be \label{bik_c}
c= \sqrt{\frac{s}{2}}(1-2a).
\ee
To consider only positive speeds, we will restrain - without loss of generality - our analysis to the range
$0<a<\frac{1}{2}$.
In this case, the propagation of the front translates as an invasion of the state $0$ by the state $1$.

At each time, the front is centered around $x=ct$, this means $u(ct, t) = \tilde{U}_{bi}(0) = \frac{1}{2}$.
We can then define the width $w$ of a given front by the relation
\begin{equation*}
\tilde{U}_{bi}(-w) = \frac{1}{1+e^{-1}} \approx 0.73
\end{equation*}
or equivalently
$\tilde{U}_{bi}(w) = \frac{1}{1+e} \approx 0.27$.
With that definition we have here
\be \label{bik_w}
w=\sqrt{\frac{2}{s}}
\ee

In the monostable case, the situation is somewhat different.
There are traveling wave solutions for a continuum of speeds $c \geq 2\sqrt{s}$,
and here again traveling wave solutions are fronts going from $1$ at $-\infty$ to $0$ at $+\infty$.
It is known that asymptotically,
a great number of solutions converge to the front of minimal speed $c = 2\sqrt{s}$.
That is notably the case when $u_0$ is compactly supported
\cite{roquejoffre1994}.
Only one family of exact solutions is known \cite{ablowitz},
they take the form
\be \label{exactmono}
u(x,t) = \tilde{U}_{mono}(x-ct) = \frac{1}{ \left(1+ K~\exp \left(\sqrt{\frac{s}{6}} (x-ct) 
\right) \right)^2}
\ee
where $K$ is a constant and the speed is given by
\be \label{mok_c}
c = 5\sqrt{\frac{s}{6}}.
\ee
We will use this solution with $K = \sqrt{2}-1$ to ensure that $\tilde{U}_{mono}(0) = \frac{1}{2}$.
Here we define the width $w$ of a given front by the relation
\begin{equation*}
\tilde{U}_{mono}(-w) = \frac{1}{(1+(\sqrt{2}-1)e^{-1})^2} \approx 0.75
\end{equation*}
or equivalently
$\tilde{U}_{mono}(w) = \frac{1}{(1+(\sqrt{2}-1)e)^2} \approx 0.22$.
With that definition we have
\be \label{mok_w}
w=\sqrt{\frac{6}{s}}
\ee
Some differences between the bistable case and the monostable one are better understood
when we see waves in the bisable case as being pushed by the bulk of the population distribution
and waves in the monostable case wave as being pulled by the leading edge of the distribution
\cite{garnier2012}.

\section{Collective variables}

In the model we are interested in, the reactivity $s$ is allowed to be space-dependent.
When the variations of $s$ are small, it is reasonable to expect that the solutions of \eqref{rd}
will remain close to the solutions in the homogeneous case.
This approach is sometimes called the use of collective variables.
First, we can expect that the front remains close to its original
profile but will move with a modulation of its speed:
\begin{equation*}
u(x,t) \approx \tilde{U}(x-c(t)t)
\end{equation*}
Actually, 
the numerical simulations indicate that the solution behavior is better captured through
a modulation of its center $x_0$ and width $w$:
\begin{equation*}
u(x,t) \approx \tilde{U}\left(\frac{x-x_0(t)}{w(t)}\right)
\end{equation*}
When investigating the evolution of a compactly supported initial condition,
for instance growing from near $0$ values,
it can also be useful to allow for a modulation of the amplitude
in the form of the following ansatz:
\begin{equation*}
u(x,t) \approx A(t) \tilde{U}\left(\frac{x-x_0(t)}{w(t)}\right)
\end{equation*}
This kind of approach is used for instance in \cite{bcp14},
where evolution equations for the collective variables are obtained through balance laws,
just like in \cite{cs1}.
As our goal is to investigate the dynamics of established fronts,
and not the evolution from an initial condition to a generalized traveling front,
we will make no use here of this third collective variable.

That being said, in the context of this paper,
we will assume that at order zero we have
\be\label{front_ansatz}
u(x,t) = U(x, x_0(t), w(t)) := (\tilde{U} \circ z) (x, x_0(t), w(t))
\ee
where $\tilde{U}$ is a given profile and we define
$z (x, x_0(t), w(t)) = \frac{x-x_0(t)}{w(t)}$.
We can compute the time derivative and the second order space derivative
\begin{equation*}
\left\{
\begin{aligned}
u_t    &= \dot{x_0} \frac{\pa U}{\pa x_0} + \dot{w} \frac{\pa U}{\pa w} \\
u_{xx} &= \frac{\pa^2 U}{\pa x^2}
\end{aligned}
\right.
\end{equation*}
The next step is to obtain time evolution equations for the collective variables
$x_0$ and $w$.
To that end, we follow the procedure exposed in \cite{cs1} and \cite{susanto}.
After the analysis is carried over (details of the computations are given in appendix \ref{calculs}),
we obtain the system of ordinary differential equations:
\begin{subequations}
\label{dx0dw_abstract}
\begin{align}
\dot{x_0} &= \frac{\alpha_1}{w}
+ \alpha_2 w K_0(x_0, w, s, R)
+ \alpha_3 w K_1(x_0, w, s, R) \label{dx0dw_abstract_eqn1} \\
\dot{w} &= \frac{\alpha_4}{w}
+ \alpha_5 w K_0(x_0, w, s, R)
+ \alpha_6 w K_1(x_0, w, s, R) \label{dx0dw_abstract_eqn2}
\end{align}
\end{subequations}
where the $\alpha_i$ are numbers and the integrals $K_n$ are given by
\be\label{Kn}
K_n(x_0, w, s, R) = \int_{-\infty}^{\infty} z^n s(wz + x_0)~R(\tilde{U}(z)) \tilde{U}'(z) ~dz.
\ee
The integrals $K_n$ are the main driver of the front time evolution, as we will see in the next sections.

\subsection{\label{subsec:eqns-cllc-variables}Time-evolution equations for the collective variables}
In the case of a bistable reaction term $R=R_b$,
we are always able to compute the integrals
because there is unicity of the profile
\be \label{bistable_front}
\tilde{U}(z)=\frac{1}{1 + \mathrm{exp}(z)}
\ee
and the equations \eqref{dx0dw_abstract} reduce to:
\begin{subequations}
\label{dx0dw_bi}
  \begin{align}
  \dot{x_0} &= -6 w K_0(x_0, w, s, R_b) \label{dx0dw_bi_eqn1} ,\\
  \dot{w}   &= \frac{3}{2 (\pi^2-6) w} -\frac{18}{\pi^2-6} w K_1(x_0, w, s, R_b) \label{dx0dw_bi_eqn2},
  \end{align}
\end{subequations}
These equations were obtained by Dawes and Susanto \cite{susanto}.

In the case of a monostable reaction term $R=R_m$,
we are able to explicitly compute the integrals
only when the profile $\tilde{U}$ is known.
This is especially the case if we consider the simplest following ansatz
\be \label{Ablowitz_front}
\tilde{U}(z)=\frac{1}{(1 + (\sqrt{2}-1)\mathrm{exp}(z))^2}
\ee
which is an exact solution of \eqref{rd} when $s$ is constant.
Then \eqref{dx0dw_abstract} give a system silimar to \eqref{dx0dw_bi} though a bit more intricate:
\begin{subequations}
\label{dx0dw_mo}
  \begin{align}
  \dot{x_0} &= \frac{\alpha_1}{w}
  + \alpha_2 w K_0(x_0, w, s, R_m)
  + \alpha_3 w K_1(x_0, w, s, R_m) \label{dx0dw_mo_eqn1} \\
  \dot{w} &= \frac{\alpha_4}{w}
  + \alpha_5 w K_0(x_0, w, s, R_m)
  + \alpha_6 w K_1(x_0, w, s, R_m) , \label{dx0dw_mo_eqn2}
    \end{align}
\end{subequations}
and the constants are
(see appendix \ref{integrals} for the definitions of the $I_k$):
\begin{align*}
\alpha_1 &= \frac{90 I_1}{85 - 12\pi^2} \approx -0.026 &
\alpha_2 &= \frac{900 I_2}{85 - 12\pi^2} \approx -5.0 \\
\alpha_3 &= \frac{900 I_1}{12\pi^2 - 85} \approx 0.26 &
\alpha_4 &= \frac{18}{12\pi^2 - 85} \approx 0.54 \\
\alpha_5 &= \frac{900 I_1}{12\pi^2 - 85} \approx 0.26 &
\alpha_6 &= \frac{180}{85 - 12\pi^2} \approx -5.4 \\
\end{align*}

\subsection{Defect wide compared to the front width}
For a defect that varies on a scale longer than $w$, $s(x_0+wz)\approx s(x_0)$,
so that $s$ goes out of the integral, allowing us to take the computation
one step further.
We obtain for the bistable case
\begin{equation} \label{bi_adia}
\left\{
\begin{aligned}
\dot{x_0} &= \frac{1-2a}{2} w s(x_0) \\
\dot{w}   &= \frac{3}{2 (\pi^2-6) w} - \frac{3}{4(\pi^2-6)} w s(x_0)
\end{aligned} 
\right.
\end{equation}
and for the monostable case
\be \label{mono_adia}
\left\{
\begin{aligned}
\left( \frac{12\pi^2-85}{90} \right) \dot{x_0}
&= \frac{5 - 6 \ln(1+\sqrt{2})}{30 w}
\\
&+ \frac{30 \pi^2 - 220 + 9 \ln(1+\sqrt{2})}{270} w s(x_0)
\\
\left( \frac{12\pi^2-85}{90} \right) \dot{w}
&= \frac{1}{5 w}
- \frac{1}{30} w s(x_0)
\\
\end{aligned} 
\right.
\ee
These equations justify the notion of a local speed and width of the front.
When $s(x)=s$ is constant, we recover the speeds and widths of
the kinks respectively \eqref{bik_c} \eqref{mok_c} and \eqref{bik_w} \eqref{mok_w}
in the homogeneous situation.

For both the bistable and monostable cases, the right hand side of the 
$\dot{x_0}$ equation is always positive so that no pinning of the front occurs for wide defects.
The numerical solutions match well the predictions given by \eqref{bi_adia} and \eqref{mono_adia},
see \cite{cs1,susanto} for the bistable case and \cite{bcp14} for the monostable case.

\subsection {Defect narrow compared to the front width}

We now turn our attention to defects that are narrow in comparison to the width of the incident front.
Namely, we consider here the case of a Heaviside function.
The defect function $s(x)$ can be written
\be\label{s_more_reac}
\boxed{
s(x) = s_0 + s_1 H(x).
}
\ee
The integrals $K_i(x_0, w, s, R)$ can be broken down into two parts:
\begin{equation*}
K_i(x_0, w, s, R) = s_0 \int_{-\infty}^{+\infty} + s_1 \int_{\frac{-x_0}{w}}^{+\infty}
\end{equation*}
We introduce the expressions
\begin{subequations}
\label{Ki_left_right}
\begin{align}
K_i^l(u, R) &= \int_{-\infty}^{u} z^i R(\tilde{U}(z)) \tilde{U}'(z) ~dz
\label{Ki_left}
\\
K_i^r(u, R) &= \int_{u}^{+\infty} z^i R(\tilde{U}(z)) \tilde{U}'(z) ~dz
\label{Ki_right}
\end{align}
\end{subequations}
with the exponents $l$ for left and $r$ for right. Then $K_i(x_0, w, s, R)$
for the defect \eqref{s_more_reac} can be written as
\be
K_i(x_0, w, s, R) = s_0 K_i (x_0, w, 1, R)
+ s_1 K_i^r \left(\frac{-x_0}{w}, R \right) 
\ee
Then the system \eqref{dx0dw_abstract} reads
\begin{subequations}
\label{dx0dw}
\begin{align}
\dot{x_0} = \frac{\alpha_1}{w}
&+ w s_0 \left( \alpha_2 K_0 (x_0, w, 1, R)
               + \alpha_3 K_1 (x_0, w, 1, R) \right)
\notag \\
&+ w s_1 \left( \alpha_2 K_0^r \left( \frac{-x_0}{w}, R \right)
               + \alpha_3 K_1^r \left( \frac{-x_0}{w}, R \right) \right)
\\
\dot{w} = \frac{\alpha_4}{w}
&+ w s_0 \left( \alpha_5 K_0 (x_0, w, 1, R)
               + \alpha_6 K_1 (x_0, w, 1, R) \right)
\notag \\
&+ w s_1 \left( \alpha_5 K_0^r \left( \frac{-x_0}{w}, R \right)
               + \alpha_6 K_1^r \left( \frac{-x_0}{w}, R \right) \right)
\end{align}
\end{subequations}
Up to here, everything is reaction-agnostic.
We can, as we did earlier in subsection \ref{subsec:eqns-cllc-variables}, compute the different terms in this system.
It has been done before for the bistable case \cite{susanto},
and the constants for the monostable one are given here in appendix \ref{integrals}.
We write them here for the sake of completeness, first in the bistable case:
\begin{subequations}
\begin{align}
\dot{x_0} &=
\frac{1-2a}{2} w s_0
- 6 w s_1 K_0^r \left( \frac{-x_0}{w}, R \right)
\\
\dot{w} &= \frac{3}{2(\pi^2-6)w}
+ \frac{3}{4(6-\pi^2)} w s_0
\\
&+ \frac{18}{6-\pi^2} w s_1 K_1^r \left( \frac{-x_0}{w}, R \right)
\end{align}
\end{subequations}

and second in the monostable case:
\begin{subequations}
\begin{align}
\dot{x_0} &=
- \frac{0.026}{w}
+ 0.83 w s_0
\\
&- 5 w s_1 K_0^r \left( \frac{-x_0}{w}, R \right)
+ 0.26 w s_1 K_1^r \left( \frac{-x_0}{w}, R \right)
\\
\dot{w} &= \frac{0.54}{w}
- 0.09 w s_0
\\
&+ 0.26 w s_1 K_0^r \left( \frac{-x_0}{w}, R \right)
- 5.4 w s_1 K_1^r \left( \frac{-x_0}{w}, R \right)
\end{align}
\end{subequations}

\begin{figure}[H]
\centerline{ \epsfig{file=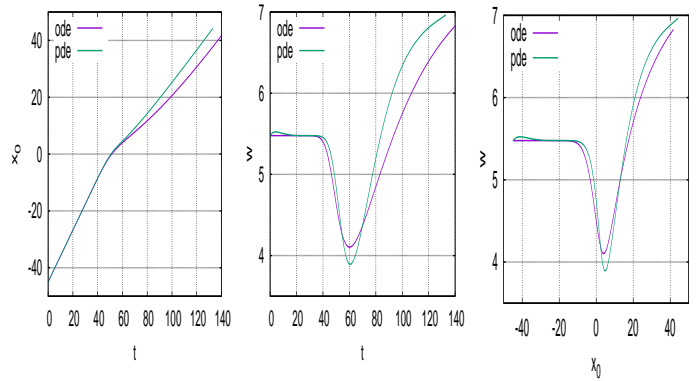,height=9 cm,width=5 cm,angle=90}}
\caption {Motion of a monostable front in a tanh like defect, 
$s = 0.2$ for negative $x$ and $s=0.1$ for positive $x$.
From left to right:
$x_0(t),~ w(t),~ w(x_0)$.}
\label{mx0w}
\end{figure}

To illustrate how well the collective variables match with the PDE
solution we consider a Heaviside defect \eqref{dx0dw_abstract}
such that 
$$s(x)= 0.2, ~~~x <0~~~~~~s(x)= 0.1, ~~~x >0 .$$
Fig. \ref{mx0w} shows $x_0(t),~ w(t),~ w(x_0)$ from left to right
for both the monostable PDE and equations (\ref{dx0dw}). The values
$x_0,w$ are obtained by fitting the PDE solution by the ansatz
(\ref{Ablowitz_front}) just as was done for the bistable solution in
\cite{cs1}. One can see the good agreement between the two solutions.

\subsection{Reaction-free zone}

We consider here a reaction-free zone.
This occurs for example in forest fires where a trench with no trees
is made to prevent fire propagation.
In combustion in a duct, this corresponds to
a region where there is no fuel.
The defect function $s(x)$ can be written
\be\label{s_no_reac}
\boxed{
s(x) = 0, ~~~0 \le x \le d,~~~s(x) = s_0 ~~~{\rm elsewhere}  .
}
\ee
With that defect,
the integrals $K_i(x_0, w, s, R)$ break down into two parts:
\begin{equation*}
K_i(x_0, w, s, R) = \int_{-\infty}^{\frac{-x_0}{w}} + \int_{\frac{d-x_0}{w}}^{+\infty}
\end{equation*}
with the naming conventions of \eqref{Ki_left_right}.
We get
\be
K_i(x_0, w, s, R) = s_0 K_i^l \left(\frac{-x_0}{w}, R \right)
+ s_0 K_i^r \left(\frac{d-x_0}{w}, R \right) 
\ee
so that the system \eqref{dx0dw_abstract} reads
\begin{subequations}
\label{dx0z}
\begin{align}
\dot{x_0} = \frac{\alpha_1}{w}
&+ w s_0 \left( \alpha_2 K_0^l \left( \frac{-x_0}{w}, R \right)
               + \alpha_3 K_1^l \left( \frac{-x_0}{w}, R \right) \right)
\notag \\
&+ w s_0 \left( \alpha_2 K_0^r \left( \frac{d-x_0}{w}, R \right)
               + \alpha_3 K_1^r \left( \frac{d-x_0}{w}, R \right) \right)
\label{dwz}
\\
\dot{w} = \frac{\alpha_4}{w}
&+ w s_0 \left( \alpha_5 K_0^l \left( \frac{-x_0}{w}, R \right)
               + \alpha_6 K_1^l \left( \frac{-x_0}{w}, R \right) \right)
\notag \\
&+ w s_0 \left( \alpha_5 K_0^r \left( \frac{d-x_0}{w}, R \right)
               + \alpha_6 K_1^r \left( \frac{d-x_0}{w}, R \right) \right)
\end{align}
\end{subequations}
Again, up to here, the computation doesn't depend on the shape of the reaction term $R$.

\section{Reaction-free zone for bistable}

In the bistable case, the major result is that the front can be stopped
for a wide enough reaction-free zone. We first illustrate this on a 2D
example and then go on to analyze it in 1D.

Consider the 2D reaction-diffusion equation
\be \label{rd2d}
u_t = u_{xx} +u_{yy}  +  s(x)~R(u) 
\ee
in a rectangular domain $|x| \le 150, ~~~|y| \le 150$ with
homogeneous Neuman boundary conditions. We assume the bistable
nonlinearity $R(u)=u(1-u)(a-u)$ with $a=0.3$. The reaction free
region is 
$$0  \le x \le 40 .$$
The initial condition is a square pulse centered at $x=-20,~~y=0$
of extension 12.
\begin{figure}[H] 
\centerline{ \epsfig{file=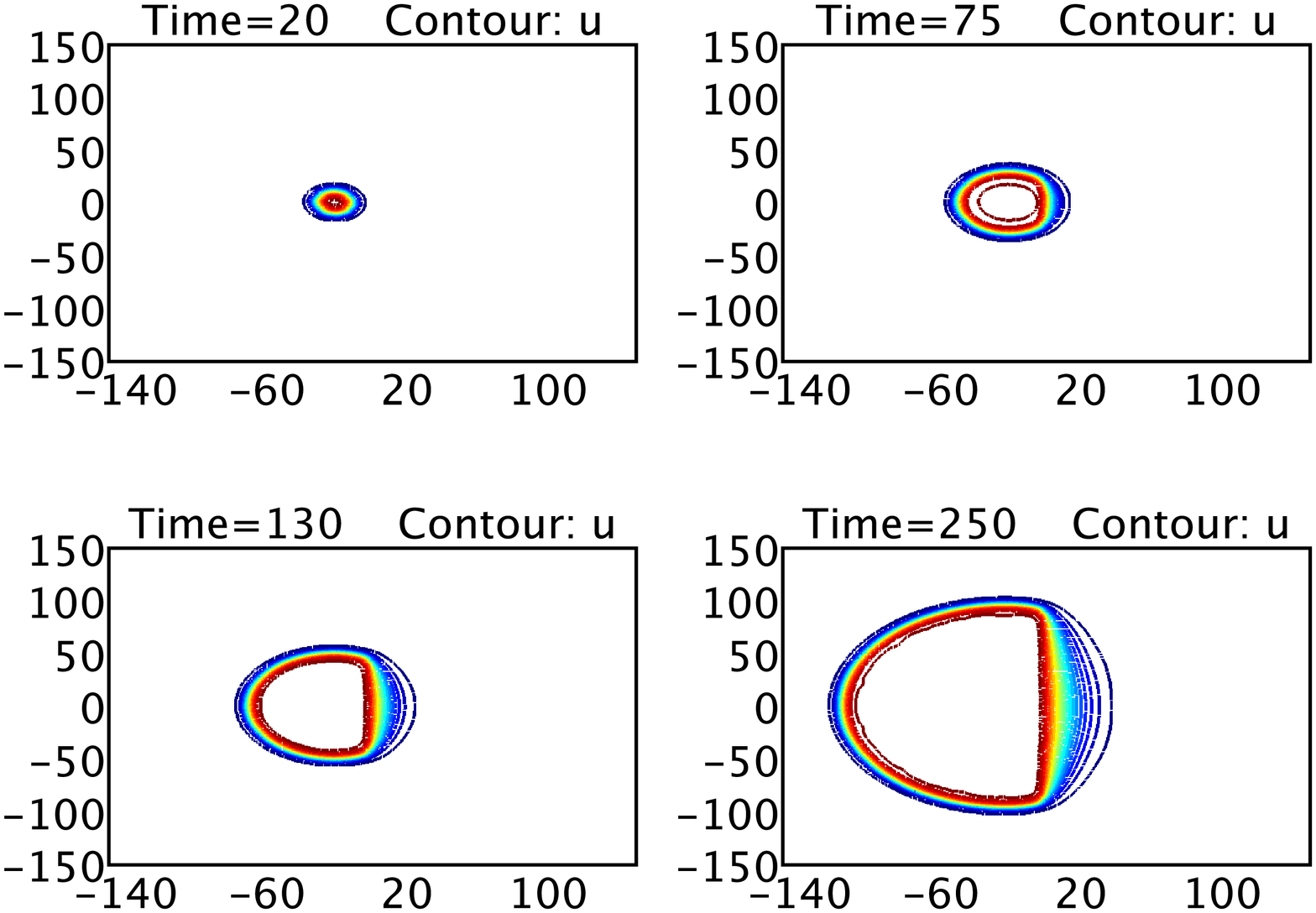,height=9 cm ,width=5 cm,angle=90}}
\caption{Motion of a two-dimensional bistable front reaching the reaction-free 
region $0\leq x \leq 40$.
The four panels show the contour lines of the solution for increasing times.
}
\label{comsolkpz}
\end{figure}
The time evolution of $u(x,y,t)$ was computed using the Comsol
software \cite{comsol}, snapshots are shown in Fig. \ref{comsolkpz} 
as contour lines ranging from $0$ (dark blue online) to $1$ (red online).
After having assumed a radial shape, the pulse hits the defect. It then
extends along it in the $y$ direction. There is not much difference
in the region $x \ge 0$ between the profiles at times $t=130$ and
$250$. This indicates that the wave cannot enter the region
$x \ge 0$.

To understand this effect quantitatively, we return to the 1D case.
The analysis is developed in the following subsections.

\subsection{\label{criterion}Criterion for front stopping}
In this section, we analyze the stopping of the front.
For simplicity of the analysis, we assume $s_0=1$ because the 
amplitude $s_0$ of the defect $s(x)$ can be scaled
out of the problem by the following change of variables
\be\label{scalings}
x= {x'\over \sqrt{s_0}},~~t= {t'\over s_0}, ~~s= s_0 s'(x').
\ee
When the front stops, $x_0$ and $w$ are stationary. 
The collective variable odes (\ref{dx0z}) yield
\begin{subequations}
\label{stationary-syst}
\begin{align}
- \frac{\alpha_1}{w} &=
w s_0 \left( \alpha_2 K_0^l \left( \frac{-x_0}{w}, R \right)
               + \alpha_3 K_1^l \left( \frac{-x_0}{w}, R \right) \right)
\notag \\
&+ w s_0 \left( \alpha_2 K_0^r \left( \frac{d-x_0}{w}, R \right)
               + \alpha_3 K_1^r \left( \frac{d-x_0}{w}, R \right) \right)
\label{stationary-syst-1}
\\
- \frac{\alpha_4}{w} &=
w s_0 \left( \alpha_5 K_0^l \left( \frac{-x_0}{w}, R \right)
               + \alpha_6 K_1^l \left( \frac{-x_0}{w}, R \right) \right)
\notag \\
&+ w s_0 \left( \alpha_5 K_0^r \left( \frac{d-x_0}{w}, R \right)
               + \alpha_6 K_1^r \left( \frac{d-x_0}{w}, R \right) \right)
\label{stationary-syst-2}
\end{align}
\end{subequations}
then
\begin{subequations}
\begin{align}
- \frac{\alpha_1 \alpha_4}{w^2 s_0} &=
\alpha_2 \alpha_4 K_0^l \left( \frac{-x_0}{w}, R \right)
               + \alpha_3 \alpha_4 K_1^l \left( \frac{-x_0}{w}, R \right)
\notag \\
&+ \alpha_2 \alpha_4 K_0^r \left( \frac{d-x_0}{w}, R \right)
               + \alpha_3 \alpha_4 K_1^r \left( \frac{d-x_0}{w}, R \right)
\\
- \frac{\alpha_1 \alpha_4}{w^2 s_0} &=
\alpha_1 \alpha_5 K_0^l \left( \frac{-x_0}{w}, R \right)
               + \alpha_1 \alpha_6 K_1^l \left( \frac{-x_0}{w}, R \right)
\notag \\
&+ \alpha_1 \alpha_5 K_0^r \left( \frac{d-x_0}{w}, R \right)
               + \alpha_1 \alpha_6 K_1^r \left( \frac{d-x_0}{w}, R \right)
\end{align}
\end{subequations}
The system implies
an equation that can be written as 
\be \label{flfr}
F^l \left( \frac{x_0}{w}, R \right)
=
F^r \left( \frac{d-x_0}{w}, R \right),
\ee
Equation \eqref{flfr} is verified for a critical distance
$$d = d_c.$$

\begin{figure}[H]
\centerline{
\epsfig{file = 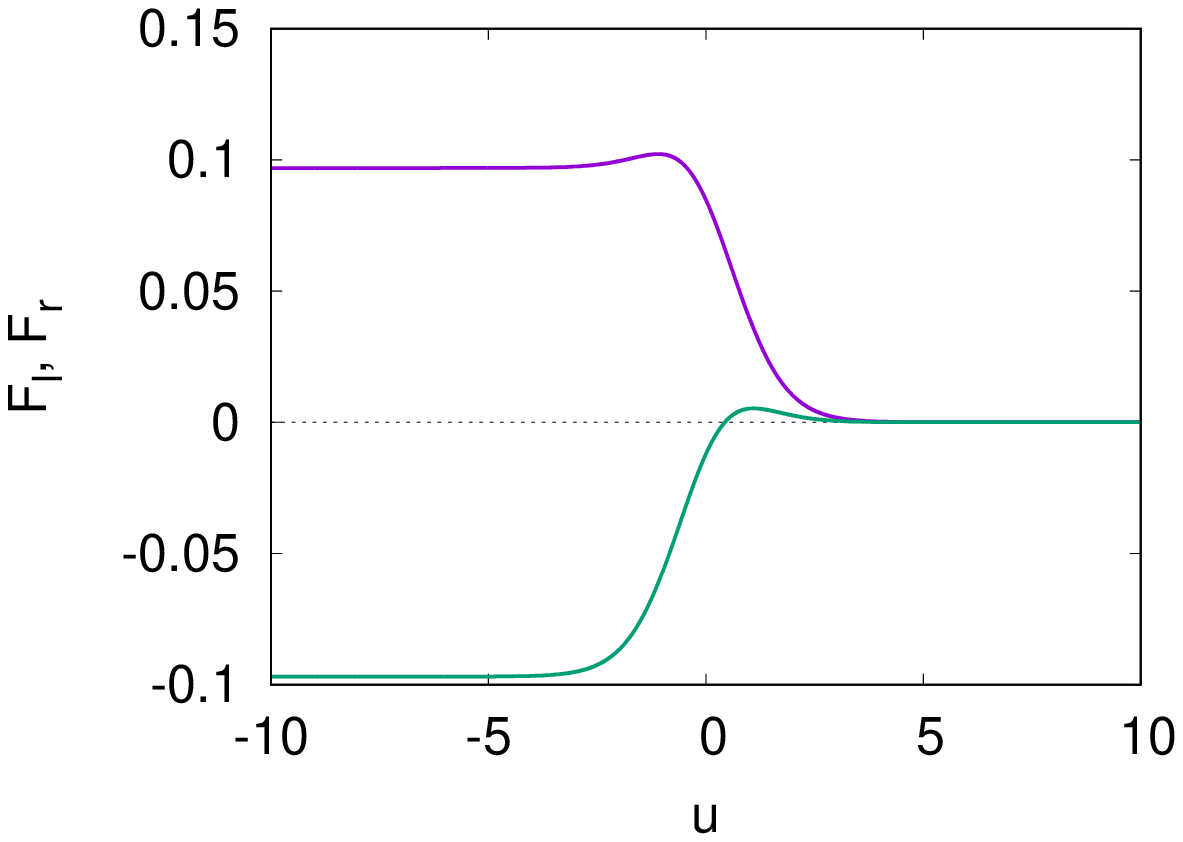,height=5cm,width=4.5 cm}
\epsfig{file = 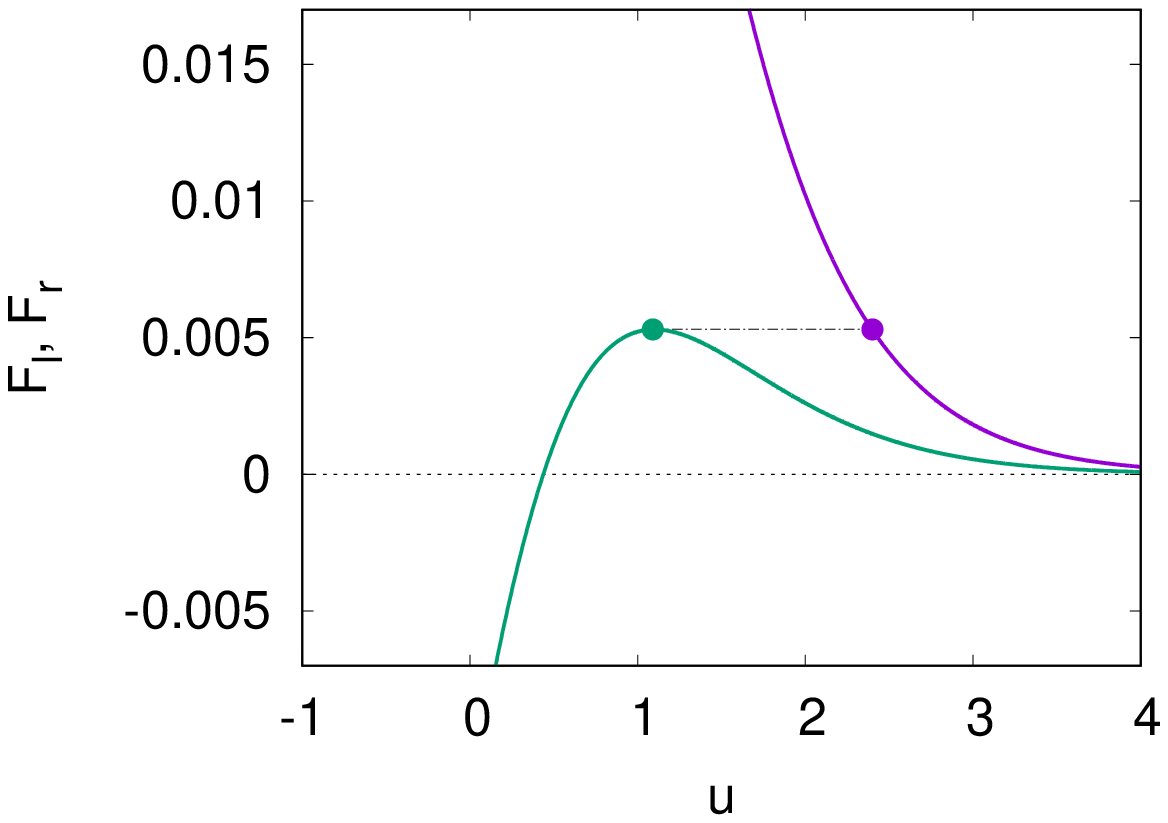,height = 5cm,width=4.5 cm}
}
\caption{Left: the functions $F^l(u, R_b)$
and $F^r(u, R_b)$.
Right: zoom on the $[-1, 4]$ region.
}
\label{sourceterms_bistable}
\end{figure}

\begin{figure}[H]
\centerline{
\epsfig{file=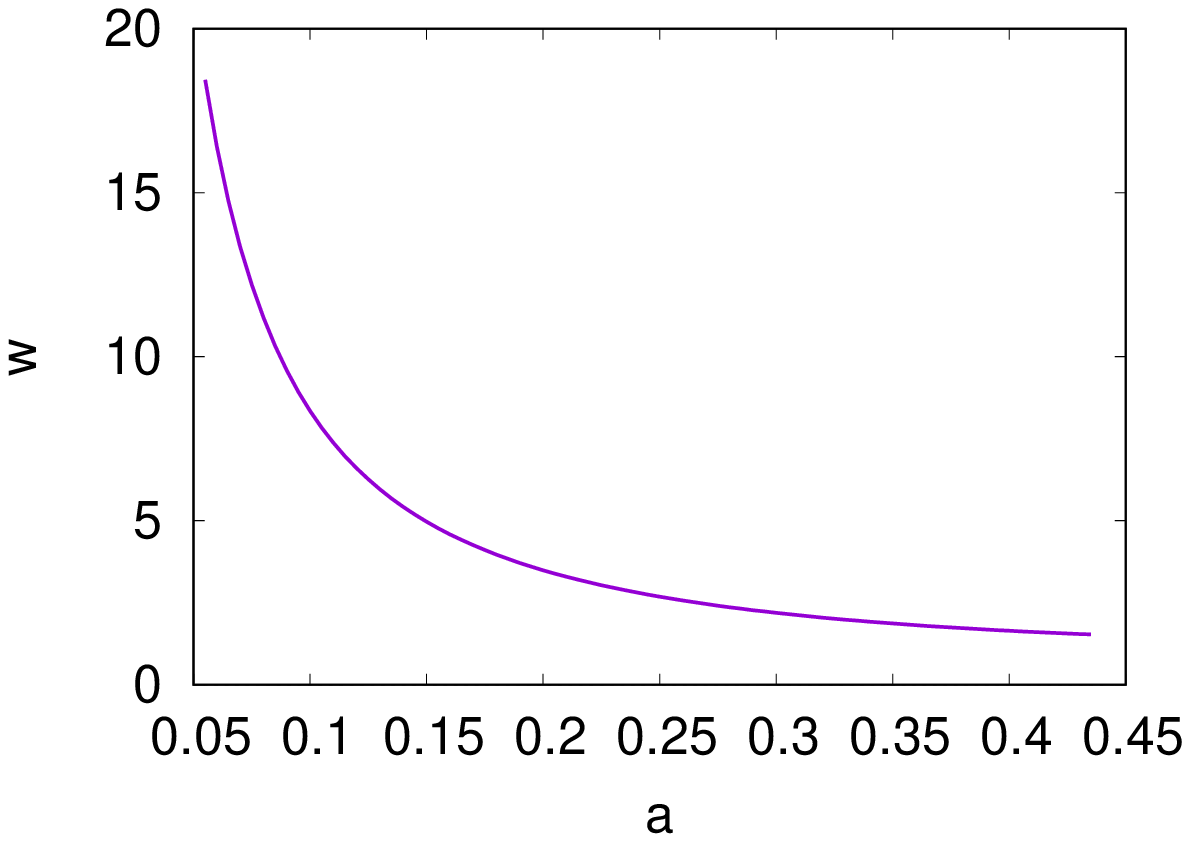,height=4cm,width= 4.5 cm,angle=0}
\epsfig{file=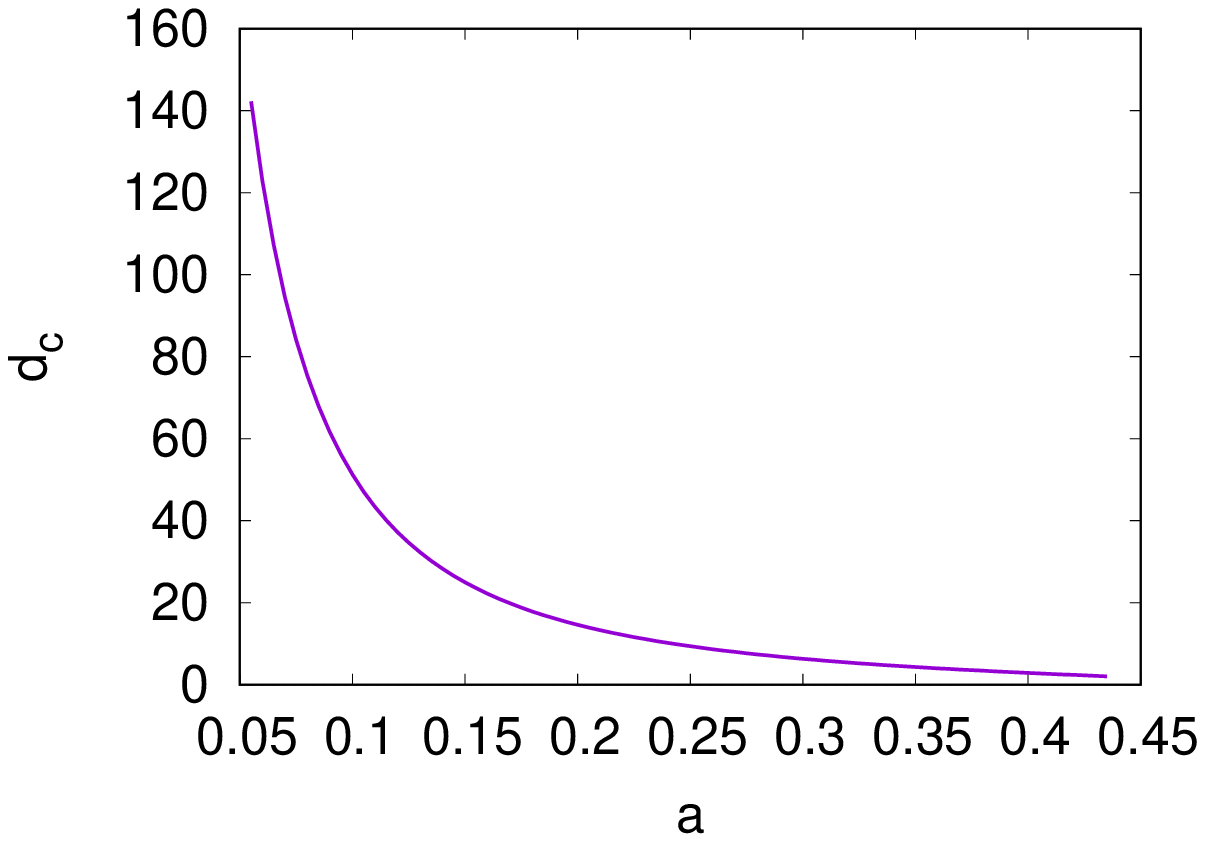,height=4cm,width= 4.5 cm,angle=0}
}
\caption{Left: width of the front when pinning occurs as a function of the parameter $a$.
Right: critical distance of the non-reaction zone as a function of the parameter $a$.
}
\label{critical_distance_and_width_pinning}
\end{figure}

The feasibility of pinning can then easily be checked by 
studying the functions $F^l$ and $F^r$; these are
plotted in Fig. \ref{sourceterms_bistable} for the bistable
nonlinearity.
For the critical distance $d_c$, we need the maximum of $F_r$ to cancel out the contribution from $F_l$.
If we define $x_r = \rm{Argmax}(F_r)$, $x_l = F_l^{-1}(\max(F_r))$,
then the following equations hold:
\begin{subequations}
\begin{align}
\frac{d_c-x_0}{w} = x_r \\
\frac{x_0}{w} = x_l
\end{align}
\end{subequations}
They give in turn $w$ using \eqref{stationary-syst-2} (a bit of caution is necessary as $\alpha_1=0$ in the bistable case)
and finally $d_c$.
As an illustration, we plot in Fig. \ref{critical_distance_and_width_pinning}
the width of the front when pinning occurs (left panel),
and the critical distance (right panel),
as functions of the parameter $a$ of the bistable reaction term.
When the parameter $a = 0.3$ for instance, we have a critical distance of $6.32$
and a width at pinning of $2.19$.
They compare very well with what we obtain through a simulation of the original PDE.

\subsection{Stationary solution when the front stops}

\begin{figure}[H]
\centerline{ \epsfig{file=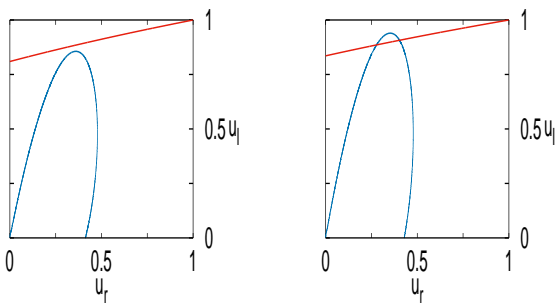,height=9 cm,width=6 cm,angle=90}}
\vskip -1cm
\caption
{Bistable case, zero reaction region $0 \leq x \leq d$. 
Plots of the functions $g(u_l,u_r)$ (curve) and $h(u_l,u_r)$ (line) in the 
$(u_l,u_r)$ plane for two different values of $d$,
$d = 6 < d_c$ (left)
and $d = 7 > d_c$ (right); the parameter is $a=0.3$.}
\label{kpzgh}
\end{figure}

Let us consider the stationary case. The problem is
\begin{align}
\label{sz1} u_{xx} + u(1-u)(u-a) &= 0 ,~~ x\le 0, \\
\label{sz2} u_{xx} &= 0 ,~~ 0 \le x\le d, \\
\label{sz3} u_{xx} + u(1-u)(u-a) &= 0 ,~~ x\ge d,
\end{align}
together with the boundary conditions
$u=1,~u_x=0,~~x\to -\infty$
and $u=0,~u_x=0,~~x\to \infty$.
The interface conditions at $x=0, d$ are continuity of $u$
and of $u_x$.
Multiplying (\ref{sz1},\ref{sz3}) by $u_x$ and integrating, we get
$${u_x^2 \over 2} + \left( -{u^4 \over 4} +{u^3 \over 3} (a+1) -{u^2 \over 2} a \right)
= C_1,~~~  x\le 0 . $$ 
$${u_x^2 \over 2} + \left( -{u^4 \over 4} +{u^3 \over 3} (a+1) -{u^2 \over 2} a \right)
= 0,  ~~~x\ge d . $$ 
The constant $C_1$ is obtained from the first expression evaluated
for $x \to -\infty$ where $u=1$ and $u_x=0$. We get
\be \label{c1} C_1 =   {1-2 a \over 12}. \ee
Inside the strip $0 \le x \le d$, the solution is
$$u = \alpha x + \beta .$$
Using the interface conditions at $x=0, d$, we get the two relations
for the unknowns $\alpha , \beta$.
\begin{align}
\label{ab1} {\alpha ^2 \over 2}
&-{1\over 4}  \beta^4
+ {1\over 3} (a+1) \beta^3
- {a\over 2} \beta^2
= C_1, \\
\label{ab2} {\alpha ^2 \over 2}
&-{1\over 4} (\alpha d + \beta)^4
+ {1\over 3} (a+1) (\alpha d + \beta)^3
- {a\over 2} (\alpha d + \beta)^2
= 0.
\end{align}
To solve this algebraic system we used a graphical method, introducing
the values $u_l= \beta$ and $u_r=\alpha d + \beta$.
In Fig. \ref{kpzgh} we plot the $0$ contour lines of the
functions $g(u_l,u_r)$ and $h(u_l,u_r)$
defined by
\begin{align}
\label{nrg} g(u_l,u_r)= {(u_l-u_r)^2 \over 2 d^2} + f(u_l)-C_1 , \\
\label{nrh} h(u_l,u_r)= {(u_l-u_r)^2 \over 2 d^2} +  f(u_r), \\
\label{nrf} f(x)= -{1 \over 4} x^4 + {a+1 \over 3 }x^3 -{a\over 2} x^2 ,
\end{align}
in the square $ (u_l,u_r) \in [0;1]\times [0;1]$.
\vskip -0.2 cm
In the left panel $d < d_c$ so that there is no solution.
In the right panel
$d > d_c$ and there are 2 solutions $(u_l,u_r)$. One of them is stable
and the other unstable, as expected from standard bifurcation theory.
For $a=0.3$ we find
$d_c \approx 6.5$ close to the value obtained from the pinning criterion
of section \ref{criterion}.

\begin{figure}[H]
\centerline{ 
\epsfig{file=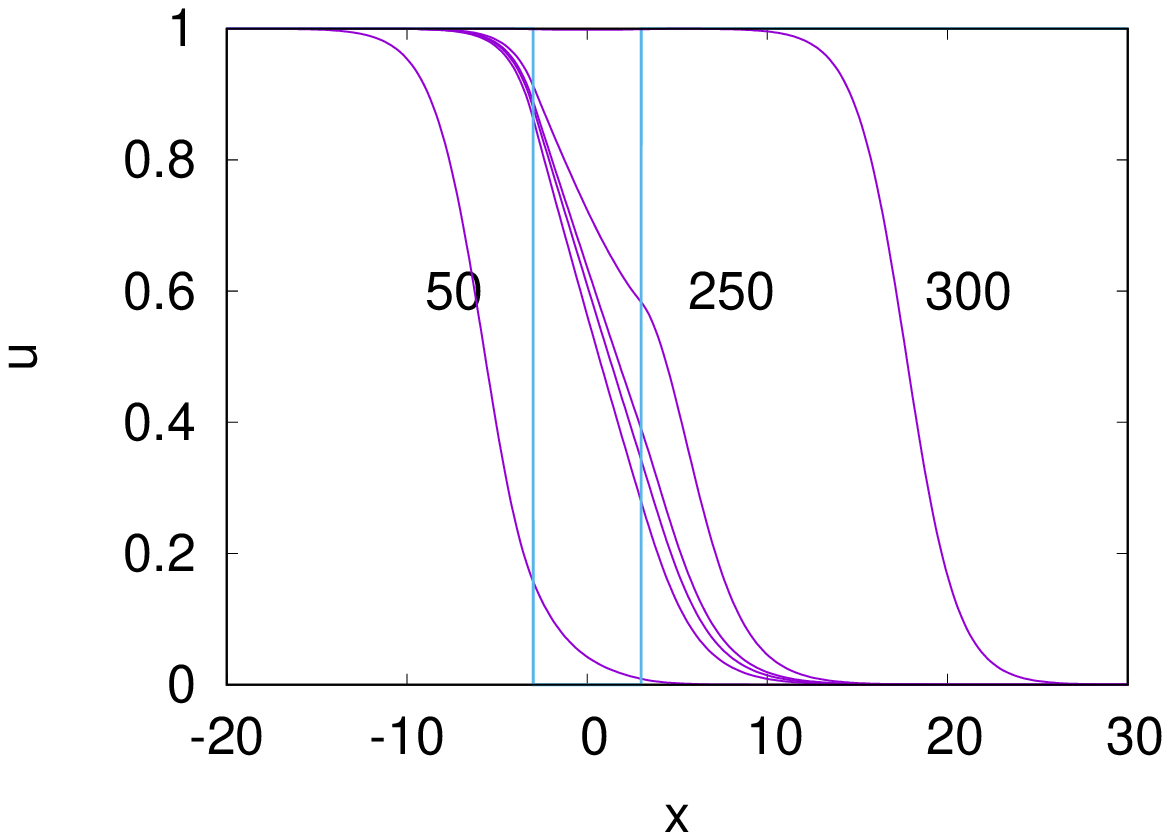,height=5 cm,width=4.5 cm,angle=0}
\epsfig{file=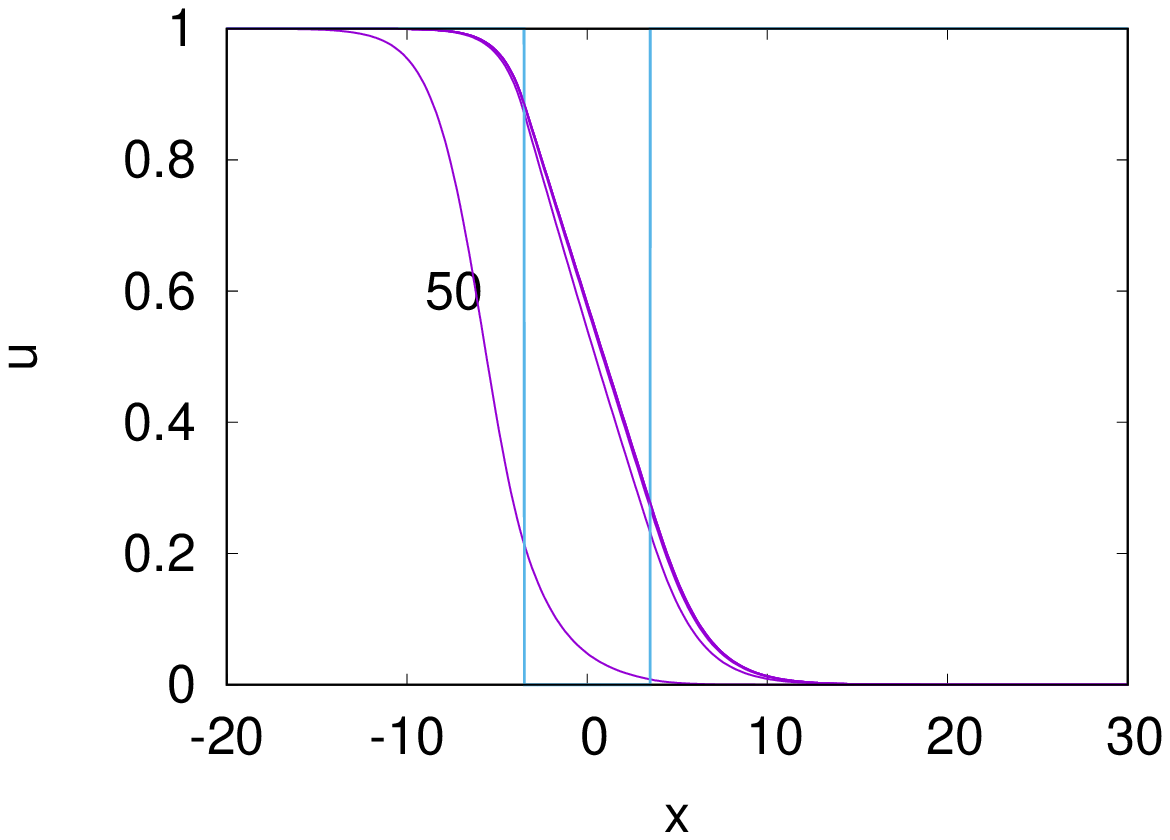,height=5 cm,width=4.5 cm,angle=0}
}
\caption
{Snapshots of
$u$ for $t=50,100,150,200,250$ and $300$ when the front crosses the 
zero reaction region, $d =6$ 
(left panel) and when the front is stopped by the zero reaction region,
$d= 7$ (right panel). }
\label{klam003}
\end{figure}

To compare with these estimates, we analyze the PDE solutions for
two different values of $d$. Fig. \ref{klam003} shows snapshots
of the solution $u(x,t)$ for different times for $d =6$
(left panel) and $d= 7$ (right panel). As expected the 
front is stopped by the zero reaction region for $d > 6.5$.
Therefore the PDE solution agrees with the analytical estimates.

\section {Reaction-free zone for monostable}

\subsection {Non existence of stationary solution }

Following the same strategy as for the bistable case, we look for a stationary
solution of the Fisher problem with a strip $[0;d]$. The formalism
is the same as above. We then get 
$${u_x^2 \over 2} + \left( {u^2 \over 2} -{u^3 \over 3} \right)
= {s \over 6},  x\le 0 . $$
$${u_x^2 \over 2} + \left( {u^2 \over 2} -{u^3 \over 3} \right)
= 0,  x\ge d .$$
There is no solution to the problem because the second expression is
always greater than 0, for $u \in [0;1]$.
This is consistent with the instability of the $u=0$ stationary state for
the monostable case.

\subsection {Appearance of secondary front}

\begin{figure}[H]
\centerline{ \epsfig{file=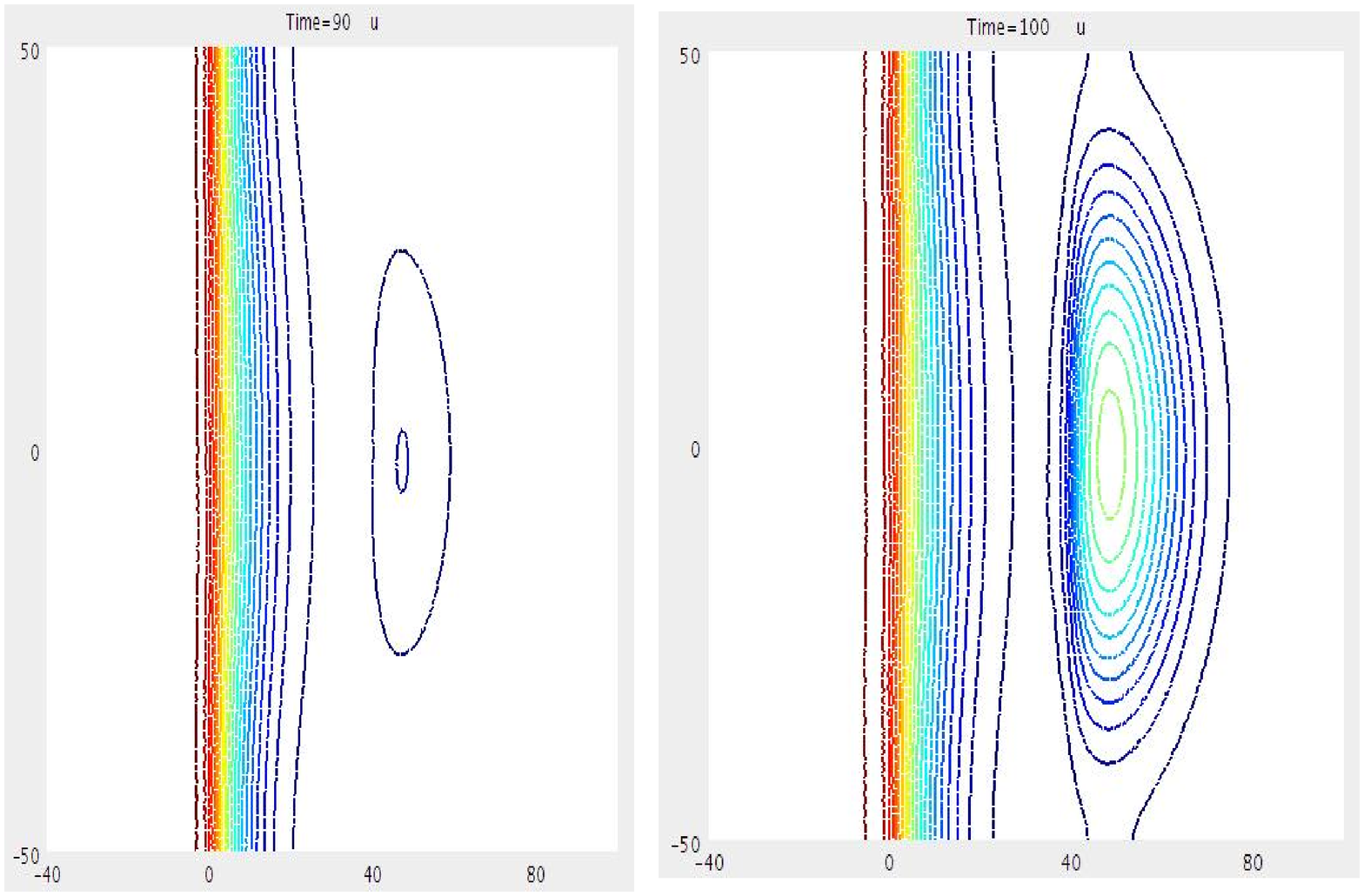,height=10 cm,width=5 cm,angle=90}}
\caption
{Monostable case, zero reaction region $0\leq x \leq 40$ and $s_0=0.3$. 
Contour lines of the solution at times $t=90$ (left) and $t=100$ (right).}
\label{comsolfish}
\end{figure}

As for the bistable model, a 2D simulation is useful to illustrate what is
happening. We consider the 2D reaction-diffusion equation (\ref{rd2d}) with
the monostable nonlinearity. Two snapshots are shown in Fig. \ref{comsolfish},
$t=90$ (left panel) and $t=100$ (right panel). For $t=90$, the front
seems to be trapped at the left interface $x=0$ of the reaction-free zone.
However, a new structure appears close to the right interface $x=40$.
The right panel of Fig. \ref{comsolfish} shows that this new structure is
a secondary pulse.

The 1D problem can be analyzed in more details. Fig. \label{fil35}
shows snapshots $u(x,t)$ of the 1D solution for seven successive 
times from $t = 27.5$ up to 192.5 for a reaction-free region $|x| \leq 35$.
The left panel shows the solution in linear scale. One can see the advancing
profile for small $u$. It reaches the right interface and becomes 
visible at $t=165$.
\begin{figure}[H]
\centerline{ \epsfig{file=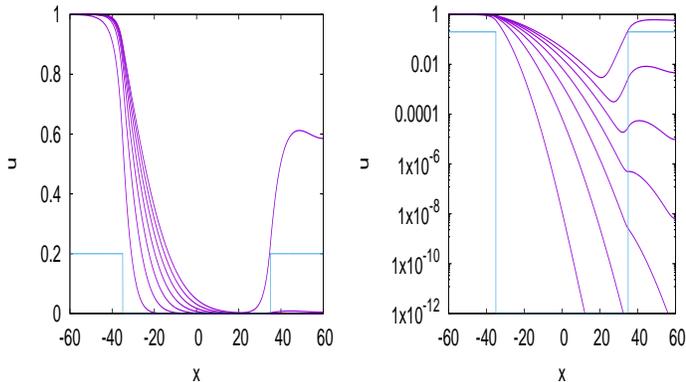,height=9 cm,width=5 cm,angle=90}}
\caption {Motion of a monostable front in a zero defect region of
extension $d=70$: snapshots of $u(x,t)$ for 
$t = 27.5, 55 , 82.5, 110, 137.5, 165$  
and 192.5 for a no reaction zone of extension $d=70$ in linear 
(left) and log (right) scales for $u$.}
\label{fil35}
\end{figure}
The increase of $u$ with time is clearer in the right panel (log plot).
There, one sees the diffusion of $u$ in the reaction-free zone
followed by its amplification for $x \geq 35$.

\begin{figure}[H]
\centerline{ \epsfig{file=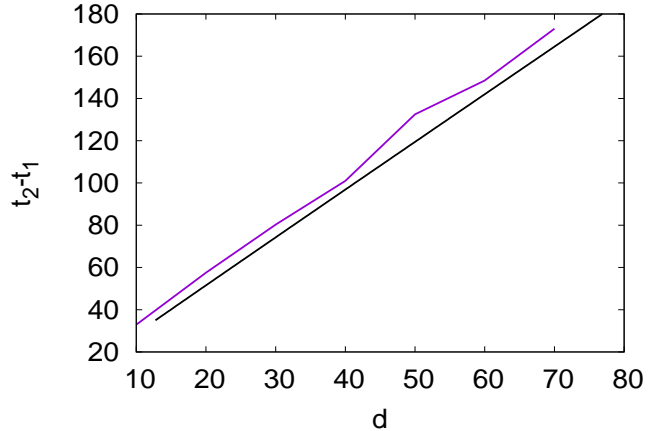,height=6 cm,width=9 cm,angle=0}}
\caption
{Time of appearance of secondary front $t_2-t_1$ for the monostable case
for different widths of the zero reaction region. The blue line 
corresponds to the numerical results and the black straight line is for the
approximation (\ref{dot}). }
\label{fish2front}
\end{figure}

The time of appearance of the secondary front can be estimated from
the solution.  Assume a zero reaction-region 
$$ 0 \leq x \leq d .$$
In Fig. \ref{fish2front} we
plot the time interval $t_2-t_1$ as a function of the half-width $d$ of the 
zero reaction region. The instant $t_1$ corresponds to the front reaching
the left edge of the strip, when its position is $x=0$. The time
$t_2$ is when the secondary pulse reaches the value $0.5$.
It is surprising that $t_2-t_1$ depends linearly on $d$. To explain this
effect, we suggest the following simple model.
In the linear (no reaction) zone, the propagation is governed by the
heat kernel giving the solution at $(x,t)$ for a source at $(y,t=0)$
\be\label{heatker}
K(x,y,t)= {1 \over \sqrt{4 \pi t}} e^{-{(x-y)^2 \over 4 t}}.\ee
Then for $x= d$, the solution of the heat equation $u_h$ is 
$u_h(d,t)={1 \over \sqrt{4 \pi t}} e^{-d^2 \over 4 t}$. 
The $u=0$ solution of the Fisher equation is unstable and any 
plane wave perturbation $e^{ikx}$ will grow as 
\be\label{dufish}
\delta u = e^{ikx +(s_0 -k^2) t}.\ee
Combining (\ref{heatker}) and (\ref{dufish}), 
the solution of the Fisher equation at $x=d$ will be
\be\label{ud} 
u(d,t)= {1 \over \sqrt{4 \pi t}} e^{-{d^2 \over 4 t} + s_0 t}, \ee
where $\xi$ is a coefficient smaller than 1 accounting for the
average of the growth rate.

We can calculate when $u(d,t)=0.5$. Using (\ref{ud})
we obtain
$$ {1 \over \sqrt{4 \pi t}} e^{-{d^2 \over 4 t} + s_0 t}=0.5 .$$
This implies
\be\label{dot} d = \sqrt{ 4 s_0 t^2 - 2 t \log (\pi t)} \ee
In Fig.  \ref{fish2front} we plotted the values of $d,t$ such that
$u(d,t)=0.5$. These follow a straight line (red) that is not far
from to the numerical results (blue). 
Despite appearances, the formula (\ref{dot}) gives
a linear dependance because of the interval in t $[20,180]$ on which
it is plotted.

\section{Conclusion}
We analyzed the stopping of a 1D reaction-diffusion front by a reaction-free
region for bistable and monostable nonlinearities.

Bistable fronts can be stopped and -using a collective variable description-
we obtained a stopping criterion linking the width of the region and
the parameter of the nonlinearity. This criterion is in good agreement
with the analysis of the time independant problem and the PDE solutions.

The monostable nonlinearity is more complex. If the front is accelerated by
the defect, the collective variables agree well with the PDE numerical solution,
otherwise a secondary pulse appears and the collective variable description is
wrong.

For a reaction-free region, we can predict the time of appearance of the
secondary pulse using a simple model based on the diffusion kernel and
the growth rate of the zero unstable state.

This reaction-free region can be generalized into a damped-reaction
region to model the action of (chemo/radio) therapy on a cancer tumor
\cite{cancer2}. The function $s(x)$ becomes
\be\label{sdampedreac}
s(x)=-s_1, ~~~0 \le x \le  d,~~~s(x)=s_0 ~~~{\rm elsewhere}  . \ee
Using the same approach as for (\ref{ud}), we find the form of
the solution in the damped-reaction region as
\be\label{ud2}
u(d,t)= {1 \over 4 \pi t} e^{{d^2 \over 4 t} + (s_0 -s_1)t}, \ee
so that we can predict the time of crossing
\be\label{dot2} d = \sqrt{ 4 (s_0-s_1) t^2 - 2 t \log (\pi t)}. \ee

The radically different behaviors between a monostable and a bistable 
front raises important questions for modelers. Does it make sense that 
a monostable front front can cross a no reaction zone of arbitrary width? 
Maybe tumor researchers should use a $ u^2(1-u)$ nonlinearity instead of 
the standard logistic. Then $u=0$ would be stable and there would be
a critical distance that the front could not cross.

\appendix

\section{\label{calculs}Derivation of the time evolution equations for the collective variables}
We multiply the equation
\begin{equation*}
\dot{x_0} \frac{\pa U}{\pa x_0} + \dot{w} \frac{\pa U}{\pa w}
= \frac{\pa^2 U}{\pa x^2}  +  s(x)~R(U(x, x_0(t), w(t)))
\end{equation*}
by the test functions $\frac{\pa U}{\pa x_0}$ and $\frac{\pa U}{\pa w}$ respectively
and we integrate over $x$ going from $-\infty$ to $+\infty$.
We obtain
\begin{equation*} 
\left\{
\begin{aligned}
&{} \dot{x_0}  \int_{-\infty}^{\infty} \left( \frac{\pa U}{\pa x_0} \right)^2 
+ \dot{w} \int_{-\infty}^{\infty} \frac{\pa U}{\pa x_0} \frac{\pa U}{\pa w} 
\\
&= \int_{-\infty}^{\infty} \frac{\pa^2 U}{\pa x^2} \frac{\pa U}{\pa x_0} 
+ \int_{-\infty}^{\infty} s(x)~R(U(x, x_0(t), w(t))) \frac{\pa U}{\pa x_0} 
\\
&{} \dot{x_0} \int_{-\infty}^{\infty} \frac{\pa U}{\pa x_0} \frac{\pa U}{\pa w} 
+ \dot{w}  \int_{-\infty}^{\infty} \left( \frac{\pa U}{\pa w} \right)^2 
\\
&= \int_{-\infty}^{\infty} \frac{\pa^2 U}{\pa x^2} \frac{\pa U}{\pa w} 
+ \int_{-\infty}^{\infty} s(x)~R(U(x, x_0(t), w(t))) \frac{\pa U}{\pa w} 
\end{aligned}
\right.
\end{equation*}
Thanks to the chain rule, we have
\begin{equation*}
\left\{
\begin{aligned}
\frac{\pa U}{\pa x} &= \frac{1}{w} \tilde{U}'(z)
\quad \mbox{and} \quad
\frac{\pa^2 U}{\pa x^2} =  \frac{1}{w^2} \tilde{U}''(z) \\
\frac{\pa U}{\pa x_0} &= -\frac{1}{w} \tilde{U}'(z) \\
\frac{\pa U}{\pa w}   &= -\frac{z}{w} \tilde{U}'(z) \\
\end{aligned}
\right.
\end{equation*}
We remark that it is convenient to make the change of variable $z = \frac{x-x_0(t)}{w(t)}$
in each integral to get (after multiplication by $w$):
\begin{equation*}
\left\{
\begin{aligned}
&{} \dot{x_0}  \int_{-\infty}^{\infty} \tilde{U}'(z)^2 ~
+ \dot{w} \int_{-\infty}^{\infty} z \tilde{U}'(z)^2 
\\
&= -\frac{1}{w} \int_{-\infty}^{\infty} \tilde{U}'(z) \tilde{U}''(z) 
- w \int_{-\infty}^{\infty} s(wz + x_0)~R(\tilde{U}(z)) \tilde{U}'(z) 
\\
&{} \dot{x_0}  \int_{-\infty}^{\infty} z \tilde{U}'(z)^2 
+ \dot{w} \int_{-\infty}^{\infty} z^2 \tilde{U}'(z)^2 
\\
&= -\frac{1}{w} \int_{-\infty}^{\infty} z \tilde{U}'(z) \tilde{U}''(z) 
- w \int_{-\infty}^{\infty} z s(wz + x_0)~R(\tilde{U}(z)) \tilde{U}'(z) 
\\
\end{aligned}
\right.
\end{equation*}
The system is written in abstract form:
\begin{equation*} 
\left\{
\begin{aligned}
I_0 \dot{x_0}
+ I_1 \dot{w}
&= -\frac{J_0}{w}
- w K_0(x_0, w, s, R)
\\
I_1 \dot{x_0}
+ I_2 \dot{w}
&= -\frac{J_1}{w}
- w K_1(x_0, w, s, R)
\\
\end{aligned}
\right.
\end{equation*}
where
\begin{equation*}
\left\{
\begin{aligned}
I_n &= \int_{-\infty}^{\infty} z^n \tilde{U}'(z)^2 ~dz \\
J_n &= \int_{-\infty}^{\infty} z^n \tilde{U}'(z) \tilde{U}''(z) ~dz \\
K_n(x_0, w, s, R) &= \int_{-\infty}^{\infty} z^n s(wz + x_0)~R(\tilde{U}(z)) \tilde{U}'(z) ~dz
\end{aligned}
\right.
\end{equation*}
The integrals $I_n$ and $J_n$ are numbers, they only depend on the profile $\tilde{U}$.
In the cases that we consider, the determinant of the system is non zero and
we get the following equations governing $\dot{x_0}$ and $\dot{w}$:
\begin{equation*}
\left\{
\begin{aligned}
  \dot{x_0} &= \frac{\alpha_1}{w}
  + \alpha_2 w K_0(x_0, w, s, R)
  + \alpha_3 w K_1(x_0, w, s, R) \\
  \dot{w} &= \frac{\alpha_4}{w}
  + \alpha_5 w K_0(x_0, w, s, R)
  + \alpha_6 w K_1(x_0, w, s, R)
\end{aligned}
\right.
\end{equation*}
where
\begin{equation*}
\left\{
\begin{aligned}
\alpha_1 &= \frac{I_1 J_1 - I_2 J_0}{I_0 I_2 - I_1^2} \\
\alpha_2 &= \frac{I_2}{I_1^2 - I_0 I_2} \\
\alpha_3 &= \frac{I_1}{I_0 I_2 - I_1^2} \\
\alpha_4 &= \frac{I_1 J_0 - I_0 J_1}{I_0 I_2 - I_1^2} \\
\alpha_5 &= \frac{I_1}{I_0 I_2 - I_1^2} \\
\alpha_6 &= \frac{I_0}{I_1^2 - I_0 I_2} \\
\end{aligned}
\right.
\end{equation*}

\section{\label{integrals}Exact values of integrals present in the paper}
For the bistable case,
\begin{equation*}
\begin{aligned}
I_0 &= \int_{-\infty}^{\infty} \tilde{U}'(z)^2 ~dz
= \frac{1}{6} \\
I_1 &= \int_{-\infty}^{\infty} z \tilde{U}'(z)^2 ~dz
= 0 \\
I_2 &= \int_{-\infty}^{\infty} z^2 \tilde{U}'(z)^2 ~dz
= \frac{\pi^2 - 6}{18} \approx 0.21 \\
J_0 &= \int_{-\infty}^{\infty} \tilde{U}'(z) \tilde{U}''(z) ~dz
= 0 \\
J_1 &= \int_{-\infty}^{\infty} z \tilde{U}'(z) \tilde{U}''(z) ~dz
= - \frac{1}{12} \\
K_0(x_0, w, 1, R_b)
&= \int_{-\infty}^{\infty} R_b(\tilde{U}(z)) \tilde{U}'(z) ~dz
= \frac{2a - 1}{12}\\
K_1(x_0, w, 1, R_b)
&= \int_{-\infty}^{\infty} z R_b(\tilde{U}(z)) \tilde{U}'(z) ~dz
= \frac{1}{24} \\
\end{aligned}
\end{equation*}
and the constants in \eqref{dx0dw_abstract} read
\begin{align*}
\alpha_1 &= 0 &
\alpha_2 &= - 6 \\
\alpha_3 &= 0 &
\alpha_4 &= \frac{1}{12 I_2} \approx 0.39 \\
\alpha_5 &= 0 &
\alpha_6 &= - \frac{1}{I_2} \approx -4.7 \\
\end{align*}
For the monostable case,
\begin{equation*}
\begin{aligned}
I_0 &= \int_{-\infty}^{\infty} \tilde{U}'(z)^2 ~dz
= \frac{1}{5} \\
I_1 &= \int_{-\infty}^{\infty} z \tilde{U}'(z)^2 ~dz
= \frac{\ln(1+\sqrt{2})}{5} - \frac{1}{6} \\
&\approx 0.0096 \\
\end{aligned}
\end{equation*}

\begin{equation*}
\begin{aligned}
I_2 &= \int_{-\infty}^{\infty} z^2 \tilde{U}'(z)^2 ~dz \\
&= -\frac{1}{3} + \frac{\pi^2}{15} - \frac{\ln(1+\sqrt{2})}{3} + \frac{(\ln(1+\sqrt{2}))^2}{5} \\
&\approx 0.19 \\
J_0 &= \int_{-\infty}^{\infty} \tilde{U}'(z) \tilde{U}''(z) ~dz
= 0 \\
J_1 &= \int_{-\infty}^{\infty} z \tilde{U}'(z) \tilde{U}''(z) ~dz
= -\frac{1}{10} \\
K_0(x_0, w, 1, R_m)
&= \int_{-\infty}^{\infty} R_m(\tilde{U}(z)) \tilde{U}'(z) ~dz
= -\frac{1}{6} \\
K_1(x_0, w, 1, R_m)
&= \int_{-\infty}^{\infty} z R_m(\tilde{U}(z)) \tilde{U}'(z) ~dz \\
&= \frac{7}{45}-\frac{\ln(1+\sqrt{2})}{6}
\approx 0.0087 \\
\end{aligned}
\end{equation*}

and the constants in \eqref{dx0dw_abstract} read
\begin{align*}
\alpha_1 &= \frac{90 I_1}{85 - 12\pi^2} \approx -0.026 &
\alpha_2 &= \frac{900 I_2}{85 - 12\pi^2} \approx -5.0 \\
\alpha_3 &= \frac{900 I_1}{12\pi^2 - 85} \approx 0.26 &
\alpha_4 &= \frac{18}{12\pi^2 - 85} \approx 0.54 \\
\alpha_5 &= \frac{900 I_1}{12\pi^2 - 85} \approx 0.26 &
\alpha_6 &= \frac{180}{85 - 12\pi^2} \approx -5.4 \\
\end{align*}

\begin{acknowledgments}
Part of this work was performed using 
computing resources of CRIANN (Normandy, France).
\end{acknowledgments}


\begin{thebibliography}{99}

\bibitem{zeldovich} Ya.B. Zeldovich and D.A. Frank-Kamenetsky, K teorii ravnomernogo rasprostraneniya 
plameni, Dokladi Akademii Nauk SSSR, Vol. 19(9):693-697 (1938).

\bibitem{scott} A. Scott, Nonlinear science, emergence and dynamics of coherent structures,
Oxford University Press (2nd edition, 2003).

\bibitem{xin00} J. Xin,  Front propagation in heterogeneous media. SIAM Review, 42, 161–230, (2000).

\bibitem{cancer} K. R. Swanson, R. Rostomily, E. C. Alvord Jr, Predicting survival of patients with 
glioblastoma by combining a mathematical model and pre-operative MR imaging characteristics: a 
proof of principle. British J. Cancer , 98, 113–9, (2008).

\bibitem{murray} J. D. Murray,  Mathematical biology I: an introduction. 3rd ed. New York: Springer; 2002.

\bibitem{fisher} R. Fisher, The wave of advance of advantageous genes, Annals of Eugenics,
Vol. 7:355-369 (1937).

\bibitem{ablowitz} M. J. Ablowitz and A. Zeppetella, Explicit solutions of Fisher's equation for a 
special wave speed, Bulletin of Mathematical Biology, Vol. 41:835-840 (1979).

\bibitem{cs1} J.-G.~Caputo and B.~Sarels, Reaction-diffusion front crossing a local defect,
Phys. Rev. E 84, 041108 (2011).

\bibitem{susanto} J. H. P.~Dawes and H.~Susanto, Variational approximation and the use of 
collective coordinates, Phys. Rev. E 87, 063202 (2013).

\bibitem{bcp14}
J. Belmonte-Beitia, G. F. Calvo, V. M. Perez-Garcia, Effective particle methods for Fisher–Kolmogorov equations:
Theory and applications to brain tumor dynamics
Commun Nonlinear Sci Numer Simulat 19 (2014) 3267–3283.

\bibitem{cancer2} H. Mi, C. Petitjean, B. Dubray, P. Vera, and S. Ruan, 
Prediction of Lung Tumor Evolution During
Radiotherapy in Individual Patients With PET, 
IEEE transactions on medical imaging, vol. 33, 4, (2014).

\bibitem{bbc16} H. Berestycki, J. Bouhours and G. Chapuisat, "Front blocking and propagation in cylinders with varying cross section",
Calculus of Variations and Partial Differential Equations, Springer,
(2016).

\bibitem{thesebouhours} Juliette Bouhours, "Reaction diffusion equation in
heterogeneous media : persistance, propagation and effect of the geometry",
General Mathematics, Universit\'e Pierre et Marie Curie - Paris VI,
(2014). \\
https://tel.archives-ouvertes.fr/tel-01070608

\bibitem{roquejoffre1994} J.M. Roquejoffre, Convergence to Travelling Waves for Solutions of a 
Class of Semilinear Parabolic Equations, Journal of Differential Equations, Vol. 108(2):262-295,  (1994).

\bibitem{garnier2012} J. Garnier and T. Giletti and F. Hamel and L. Roques,
Inside dynamics of pulled and pushed fronts, Journal de Math\'ematiques Pures et Appliquées, 
Vol. 98(4):428-449 (2012).

\bibitem{comsol} https://www.comsol.com/





\end{thebibliography}
\end{document}